\documentclass[twocolumn]{revtex4-1}

\usepackage{mathpazo}
\usepackage[latin9]{inputenc}
\usepackage{xcolor}
\usepackage{amsmath}
\usepackage{amssymb}
\usepackage{graphicx}
\PassOptionsToPackage{normalem}{ulem}
\usepackage{ulem}
\usepackage[breaklinks=true,colorlinks=true,citecolor=blue,filecolor=black,linkcolor=red,urlcolor=blue]{hyperref}

\begin{document}

\title{Symmetry Breaking in Bose-Einstein Condensates Confined by a Funnel
Potential}

\author{Bruno M. Miranda}
\affiliation{Instituto de Física, Universidade Federal de Goiás, 74.690-900, Goiânia,
Goiás, Brazil}

\author{Mateus C. P. dos Santos}
\affiliation{Instituto de Física, Universidade Federal de Goiás, 74.690-900, Goiânia,
Goiás, Brazil}

\author{Wesley B. Cardoso}
\affiliation{Instituto de Física, Universidade Federal de Goiás, 74.690-900, Goiânia,
Goiás, Brazil}

\begin{abstract}
In this work, we consider a Bose-Einstein condensate in the self-focusing
regime, confined transversely by a funnel-like potential and axially
by a double-well potential formed by the combination of two inverted
Pöschl-Teller potentials. The system is well described by a one-dimensional
nonpolynomial Schrödinger equation, for which we analyze the symmetry
break of the wave function that describes the particle distribution
of the condensate. The symmetry break was observed for several interaction
strength values as a function of the minimum potential well. A quantum
phase diagram was obtained, in which it is possible to recognize the
three phases of the system, namely, symmetric phase (Josephson), asymmetric
phase (spontaneous symmetry breaking - SSB), and collapsed states,
i.e., those states for which the solution becomes singular, representing
unstable solutions for the system. We analyzed our symmetric and asymmetric
solutions using a real-time evolution method, in which it was possible
to confirm the stability of the results. Finally, a comparison with
the cubic nonlinear Schrödinger equation and the full Gross-Pitaevskii
equation were performed to check the accuracy of the effective equation
used here.

\end{abstract}
\maketitle

\section{Introduction}

\begin{figure*}[t]
\centering \includegraphics[width=0.9\textwidth]{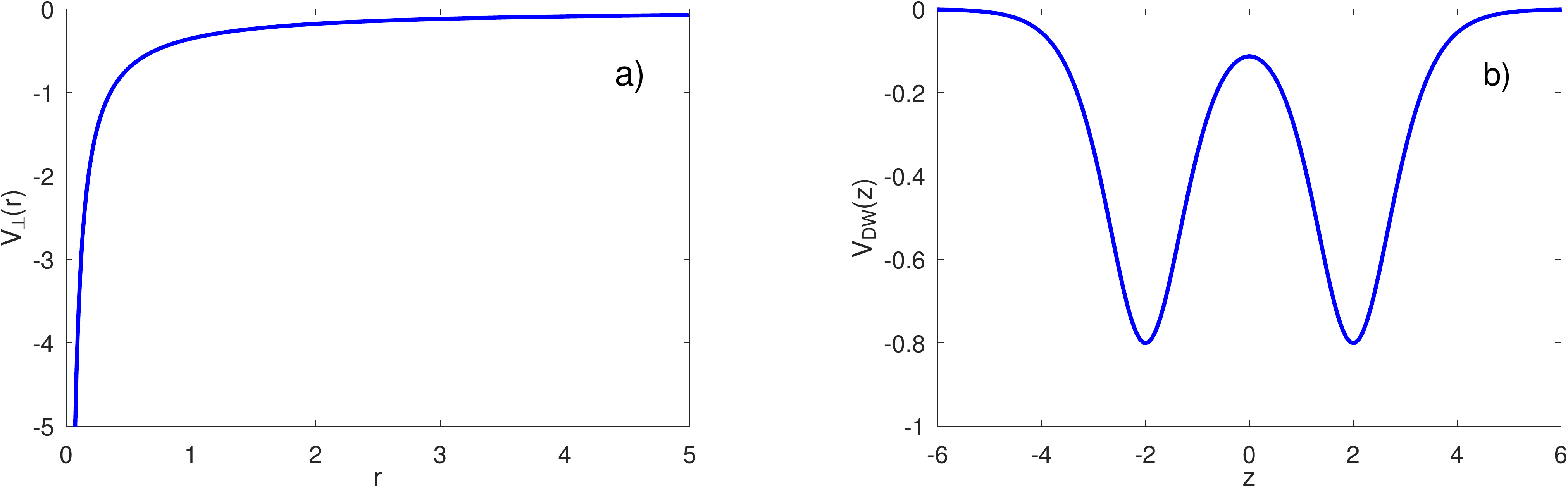} \caption{The trapping potential in the (a) transverse and (b) axial direction,
given by Eqs. (\ref{asdf}) and (\ref{Vtrans}), respectively, by
considering $V_{0}=0.8$, $z_{0}=\pm2,~a=1$ and $\varepsilon=1$.}
 \label{potenciais}
\end{figure*}

The interactions of light and matter consists of one of the most fundamental
aspects used for the realization of Bose-Einstein condensates (BECs).
Indeed, since the first realizations in diluted atomic gases of alkali
metals at ultra-low temperatures \cite{ANDERSON_SCIENCE95,DAVIS_PRL95,BRADLEY_PRL95},
several phenomena have been observed, such as formation of matter-wave
dark \cite{BURGER_PRL99} and bright \cite{ABDULLAEV_IJMPB05,CORNISH_PRL06,KHAYKOVICH_SCIENCE02}
solitons, generation of vortex states \cite{MATTHEWS_PRL99,MADISON_PRL00},
the engineering of spin--orbit-coupled BECs \cite{Lin_NATURE11},
Anderson localization of matter waves \cite{BILLY_NATURE08,ROATI_NATURE08},
quantum droplets \cite{CABRERA_SCIENCE18,CHEINEY_PRL18,SEMEGHINI_PRL18,DERRICO_PRR19},
etc.

In this context, BECs trapped by double-wells potentials have been
studied in recent decades, presenting advances from a theoretical
point of view \cite{Yuan_PLA05,Xie_MPLB15,Shchesnovich_PD04,Salasnich_MP05,MAZZARELLA_JPB09,Mazzarella_PRA10}.
In Ref. \cite{Shchesnovich_PD04}, families of bright solitons were
investigated in binary BECs, trapped by asymmetric double-well potentials.
In particular, when the scattering lengths of each component of the
BEC have opposite signs, the results (numerical and analytical) showed
stability with weak repulsive interaction. BECs trapped in double
square well and optical lattice well were also investigated in Ref.
\cite{Yuan_PLA05}.

An important phenomenon present in some systems trapped by double-well
potentials (or similar) is the spontaneous symmetry breaking (SSB),
which is a characteristic of the ground state presenting an asymmetry
with respect to the axis defined by the potential. In this sense,
in Ref. \cite{Salasnich_MP05} a coupled disk-shaped BECs, showed
symmetry breaking in a self-attractive and coupling weak regime. Even
in an almost 2D regime, bifurcation diagrams show a direct relationship
between symmetry, asymmetry and the norm of the solitons of a system
trapped in a double-channel potential (BECs or optical systems) (see
Ref. \cite{Matuszewski_PRA07} for more details). In the one-dimensional
case, single \cite{Mazzarella_PRA10} and binary \cite{MAZZARELLA_JPB09}
BECs were reported to present symmetry breaking. Furthermore, the
symmetry breaking in BECs can be induced by the nonlinearity coefficient
\cite{Mayteevarunyoo_PRA08}. In this case, the spatial dependent
nonlinearity obtained by means of the spatially inhomogeneous Feshbach
resonance, acts as a pseudo double-well potential.

In all the above-mentioned works there is the presence of some method
of dimensional reduction to describe the system by an effective 1D
or 2D equation. However, the models under consideration (BECs and
some types of optical systems) are generally described by a three-dimensional
(3D) nonlinear Schrödinger equation (NLSE), that require a lot of
computational effort to be solved numerically. On the other hand,
in Ref. \cite{SALASNICH_PRA02} the authors presented a dimensional
reduction method via a variational approach in order to obtain an
effective equation that correctly describes the longitudinal (transversal)
profile of the BEC. This technique has been extensively tested for
decades, showing great results \cite{Salasnich_PRA09,Salasnich_JPA09,Young_PRA10,Cardoso_PRE11,Salasnich_JPB12,dosSantos_PLA19,Santos_PRE20,Santos_JPB19,dosSANTOS_EPJ21}.
In the case of BECs, starting from the Gross-Pitaevskii (GP) equation
(a NLSE type), this approach allows us to derive a nonpolynomial Schrödinger
equation (NPSEs), which describes the reduced dynamics of the corresponding
physical system. Several configurations have been investigated and
recently in Refs. \cite{Santos_JPB19,dosSANTOS_EPJ21} the authors
studied a BEC confined by a radial singular potential ($\sim1/r$),
demonstrating the effectiveness of NPSE in describing the longitudinal
behavior of the ultracold gas even when confined by anharmonic potentials.
This configuration, called the \textit{funnel shape}, can be achieved
using a vapor of magnetic atoms attracted by an axial electric current.

In this paper we investigated the funnel-shaped BEC loaded into the
symmetric double-well potential (axial direction). This system is
well described by a NPSE \cite{Santos_JPB19,dosSANTOS_EPJ21}, for
which we obtain three types of ground-states: symmetric, asymmetric
and collapsed states. Symmetric solutions present densities with even
parity, i.e., the particle density is invariant by changing the sign
of its spatial coordinate. Asymmetric profiles exhibit the opposite
behavior to that shown by the symmetric case, i.e., they do not show
sign inversion symmetry of the spatial coordinate. Finally, the system
can also present the collapsed states, which have energy equal to
minus infinity \cite{Sakellari_JPB04}. We present the quantum phase
diagram of the funnel-shaped BEC and study how the characteristics
of this confinement affect the phase of the condensate. The quantum
phase of a NPSE obtained from a BEC transversely confined by a harmonic
oscillator was studied in Ref. \cite{Mazzarella_PRA10}. Differently,
here we will show results from a singular, but physical potential,
presenting a correlation between the phase and the transversal confinement.
Finally, we studied the stability of the symmetrical and asymmetrical
profiles, in addition to presenting coherent oscillations of matter
(population imbalance) produced by abrupt changes in the double-well
potential.

\section{Theoretical model}

Let us consider a system formed by a dilute quantum gas, with attractive
interaction properties, at absolute zero temperature \cite{Pethick_08}.
The system is trapped by a superposition of a funnel shape potential
in the transverse direction \cite{Santos_JPB19} and a double-well
potential in the axial direction ($z$). This double-well is formed
by the combination of two inverted Pöschl-Teller potentials, represented
by $V_{\textrm{DW}}(z)$:
\begin{eqnarray}
V_{\mathrm{DW}}(z) & = & V_{L}(z)+V_{R}(z),\label{asdf}\\
V_{L}(z) & = & -V_{0}\left[\textrm{sech}^{2}\left(\frac{z+z_{0}}{a}\right)\right],\label{asdf2}\\
V_{R}(z) & = & -V_{0}\left[\textrm{sech}^{2}\left(\frac{z-z_{0}}{a}\right)\right].\label{asdf3}
\end{eqnarray}
The funnel-shaped potential, acting in the transverse $(x,y)$ plane,
is given by
\begin{equation}
V_{\perp}(r)=-\frac{\varepsilon^{3}}{2r},\label{Vtrans}
\end{equation}
where $\varepsilon>0$ is a constant with dimension of length and
the radial coordinate is defined as $r\equiv\sqrt{x^{2}+y^{2}}$.
As an example, we display the potentials in Fig. \ref{potenciais}.

Thus, the complete trapping potential can be written as
\begin{equation}
V_{\textrm{trap}}(r,z)=-\frac{\varepsilon^{3}}{2r}+V_{\mathrm{DW}}(z).\label{Vtrap}
\end{equation}
The macroscopic wave function that describes the behavior of this
system is governed by the GP equation \cite{DALFOVO_RMP99}
\begin{equation}
i\hbar\frac{\partial}{\partial t}{\psi}(\textbf{r},t)=\left[-\frac{\hbar^{2}}{2m}\nabla^{2}+V(r,z)+Ng\left|\psi(\textbf{r},t)\right|^{2}\right]\psi(\textbf{r},t).\label{GPE3D}
\end{equation}
where $g=4\pi\hbar^{2}a_{s}/m$ is the coupling constant directly
proportional to the scattering length $a_{s}$, $N$ is the number
of particles, $m$ is the atomic mass, and $\nabla^{2}$ is the standard
Laplacian operator in three dimensions. By means of a rescale of the
variables $t\rightarrow\omega_{\perp}t$, $(x,y,z)\rightarrow(x,y,z)/a_{\perp}$,
$\psi\rightarrow\psi a_{\perp}^{3/2}$ and $V\rightarrow V/\hbar\omega$,
with $a_{\perp}=\sqrt{\hbar/m\omega_{\perp}}$, in Eq. \eqref{GPE3D},
one obtains
\begin{equation}
i\frac{\partial\psi}{\partial t}=-\frac{1}{2}\nabla^{2}\psi+V(r,z)\psi+2\pi\Gamma|\psi|^{2}\psi.\label{GPE3Dad}
\end{equation}
with $\Gamma=2a_{s}N/a_{\perp}$. The Lagrangian density that corresponds
to Eq. \eqref{GPE3Dad}, with the potential \eqref{Vtrap}, is given
by
\begin{eqnarray}
\mathcal{L} & = & \frac{i}{2}\left(\psi^{*}\frac{\partial\psi}{\partial t}-\psi\frac{\partial\psi^{*}}{\partial t}\right)-\frac{1}{2}|\nabla\psi|^{2}\nonumber \\
 & - & \left(V_{\textrm{DW}}(z)-\frac{\varepsilon^{3}}{2r}\right)|\psi|^{2}-\pi\Gamma|\psi|^{4}.\label{Lagrangean}
\end{eqnarray}
Our goal is to reduce the 3D model to an effective 1D equation. In
a didactic way, we recall here the deduction of the effective equation
shown in Ref. \cite{Santos_JPB19}, where we use the following \textit{ansatz}
\begin{equation}
\psi(r,z,t)=\exp\left(-\frac{r}{2\eta^{2}}\right)\frac{f(z,t)}{\sqrt{2\pi}\eta^{2}},\label{ansatz}
\end{equation}
where $f(z,t)$ is an axial complex wave function and $\eta=\eta(z,t)$
is the transverse length with the normalization condition 
\begin{equation}
\int_{-\infty}^{+\infty}|f(z)|^{2}dz=1,
\end{equation}
required to ensure the unitary normalization of the wave function
in 3D. Inserting the \textit{ansatz} into the Lagrangian density (Eq.
\eqref{Lagrangean}) we must to perform the integration on the transversal
plane $(x,y)$ and neglect the derivatives that include terms like
$\partial\eta/\partial z$ \cite{SALASNICH_PRA02}. Then, we obtain
the effective Lagrangian in one dimension written as
\begin{eqnarray}
L_{1D} & = & \int_{-\infty}^{\infty}\Bigg[\frac{i}{2}\left(f^{*}\frac{\partial f}{\partial t}-f\frac{\partial f^{*}}{\partial t}\right)-\frac{1}{2}\left|\frac{\partial f}{\partial z}\right|^{2}-\frac{\Gamma}{8}\frac{|f|^{4}}{\eta^{4}}\nonumber \\
 & - & \left[V_{\textrm{DW}}+\frac{1}{2\eta^{2}}\left(\frac{1}{4\eta^{2}}-\varepsilon^{3}\right)\right]|f|^{2}\Bigg]dz.
\end{eqnarray}
The following step is to use the Euler-Lagrange equation
\begin{equation}
\frac{\partial\mathcal{L}}{\partial\varphi}-\frac{\partial}{\partial t}\frac{\partial\mathcal{L}}{\partial\dot{\varphi}}-\frac{\partial}{\partial z}\frac{\partial\mathcal{L}}{\partial\varphi_{z}}=0,
\end{equation}
for both $f$ and $\eta$ terms ($\varphi\equiv f,\eta$). The results
consist of two coupled equations:
\begin{equation}
i\frac{\partial f}{\partial t}=-\frac{1}{2}\frac{\partial^{2}f}{\partial z^{2}}+V_{\textrm{DW}}(z)f+\frac{1}{2\eta^{2}}\left(\frac{1}{4\eta^{2}}-\varepsilon^{3}\right)f+\frac{\Gamma|f|^{2}}{4\eta^{4}}f,\label{feq}
\end{equation}
\begin{equation}
2\varepsilon^{3}\eta^{2}-1-\Gamma|f|^{2}=0.\label{etaeq}
\end{equation}
Inserting Eq. \eqref{etaeq} into Eq. \eqref{feq} we obtain the time-dependent
1D NPSE
\begin{equation}
i\frac{\partial f}{\partial t}=-\frac{1}{2}\frac{\partial^{2}f}{\partial z^{2}}+V_{\textrm{DW}}(z)f-\frac{\varepsilon^{6}}{2(1+\Gamma|f|^{2})^{2}}f,\label{1DNPSEcalixto}
\end{equation}
which will be the main model used in our simulations. Indeed, the
effective equation Eq. \eqref{1DNPSEcalixto} describes the dynamics
of a BEC confined in the transverse plane by a funnel-shaped potential
and axially by a double-well (Eq. \eqref{Vtrap}). Additionally, the
nonlinear regimes of strong and weak interaction can be obtained in
a simplified way through approximations. A BEC in the weak interaction
regime implies that the relation $\Gamma|f|^{2}\ll1$ is valid \cite{Pethick_08}.
Using this approximation in the above NPSE and expanding the nonpolynomial
term in terms of $\Gamma|f|^{2}$, we get the nonlinear cubic Schrödinger
equation, given by
\begin{equation}
i\frac{\partial f}{\partial t}=-\frac{1}{2}\frac{\partial^{2}f}{\partial z^{2}}+V(z)f-\varepsilon^{6}\left(\frac{1}{2}-\Gamma|f|^{2}\right)f.\label{cubicNLSE}
\end{equation}

Next, by considering a system with repulsive interatomic interaction
and in a strong interaction regime, we can make use of the well-known
Thomas-Fermi (TF) approximation \cite{Pethick_08}, which neglects
the contribution of spatial derivatives, i.e., the contribution of
kinetic energy is much smaller when compared to the interaction between
the particles. In order to obtain a one-dimensional equation for this
regime, we must substitute into the Eq. \eqref{GPE3D} a normalized
wave function, given by \cite{Santos_JPB19}
\begin{equation}
\psi(\textbf{r},t)=e^{-i\mu t}\frac{f(z)}{\sqrt{(2\pi)}}e^{-r/2},\label{psithomas}
\end{equation}
which corresponds to the initial \textit{ansatz}, Eq. \eqref{ansatz},
with $\eta=1$. Carrying out the integrations on $x$ and $y$, we
get the following result for the density distribution in $z$-direction
\begin{eqnarray}
|f(z)|^{2} & = & \begin{cases}
\dfrac{9}{4\Gamma}\left[4\mu-4V(z)+\varepsilon^{3}\right], & \;\text{for}\ \mu+\varepsilon^{3}/4>V(z)\\
0, & \;\text{for}\ \mu+\varepsilon^{3}/4\leq V(z),
\end{cases}\nonumber \\
\label{thomas}
\end{eqnarray}
which corresponds to the TF solution. In the next section we discuss
the physical effects of the inclusion of the double-well potential
in $z-$coordinate on the emergence of broken symmetry states.

\begin{figure}[tb]
\centering \includegraphics[width=0.9\columnwidth]{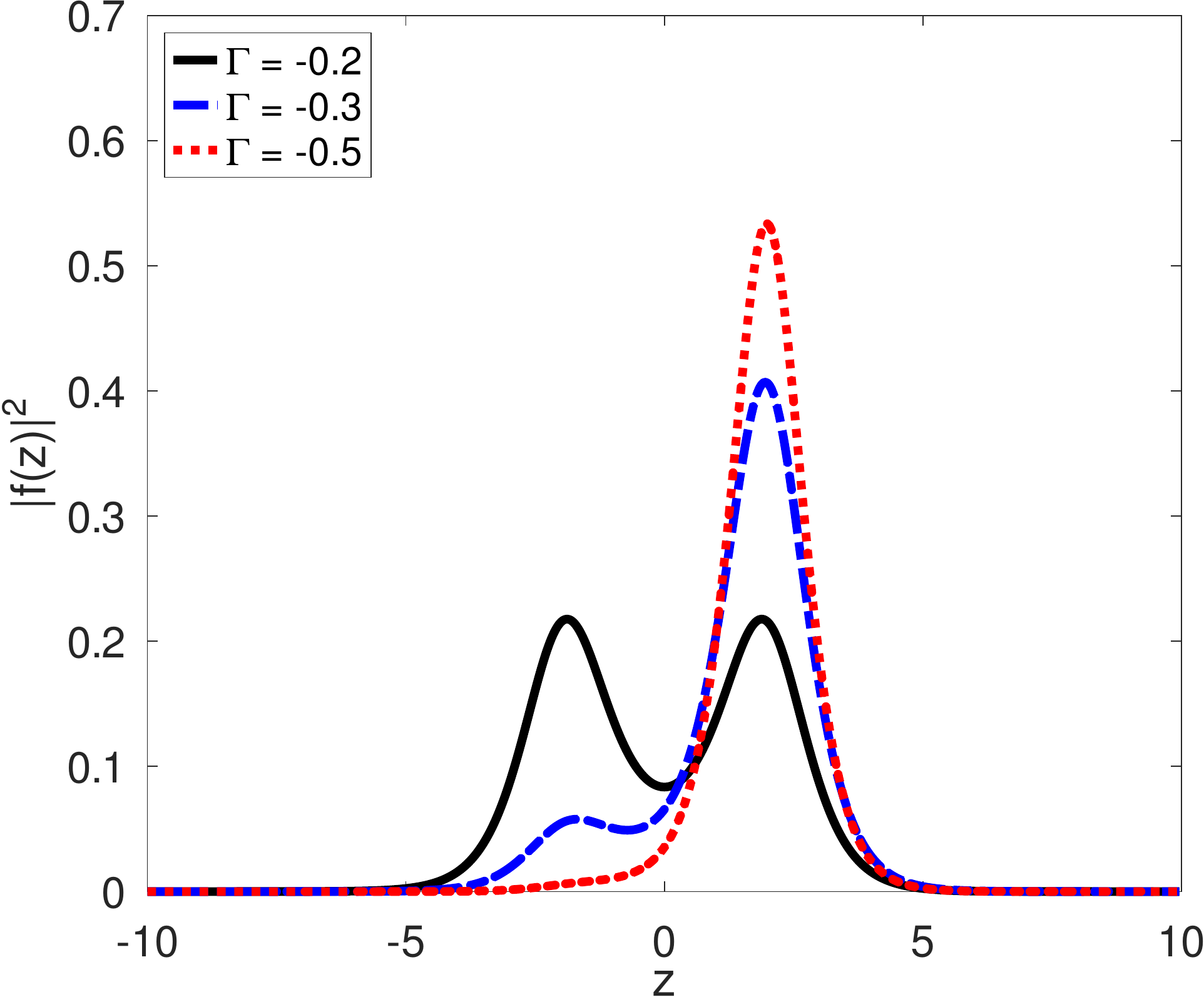} \caption{Axial probability density, $|f(z)|^{2}$, of the ground state for
different values of the interaction strength ($\Gamma$), considering
an attractive BEC trapped by a funnel potential in the transverse
direction and a double-well potential in the axial direction.}
 \label{SSBfunil}
\end{figure}

\begin{figure*}[tb]
\centering \includegraphics[width=0.9\textwidth]{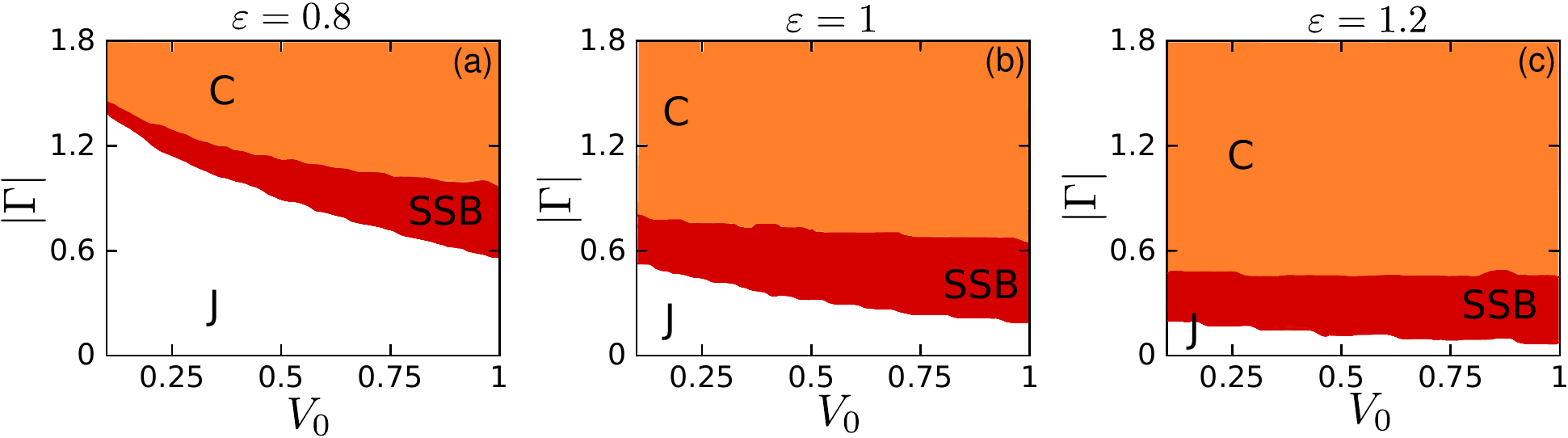} \caption{Quantum phase diagram of interaction strength $\Gamma$ vs the double-well
depth $V_{0}$ for the NPSE. Three different values of the strength
of the funnel-shaped potential are used here, i.e., (a) $\varepsilon=0.8$,
(b) $\varepsilon=1.0$, and (c) $\varepsilon=1.2$. The different
regions are represented by Josephson phase (J), spontaneous symmetry
breaking (SSB), and collapsed states (C), the latter representing
the regions whose numerical results present singularities.}
 \label{diagramanpse}
\end{figure*}

\section{Results}

First, we present the results of the imaginary-time evolution of the
1D NPSE \eqref{1DNPSEcalixto}. Indeed, by means of imaginary-time
evolution one can obtain the ground states of the system under consideration.
To this end, we have employed a second-order \textit{split step} method
to solve this equation numerically. As an example, in Fig. \ref{SSBfunil}
we plot the axial probability density of the wave function for three
different values of the interaction strength ($\Gamma$) in order
to elucidate the presence of SSB in the system. Note that for $\Gamma=-0,2$
the solution is symmetrical with respect to the double-well center,
characterizing a state known as the Josephson phase (J). For $\Gamma<-0.2$
the metastable state solution becomes asymmetric and the phenomenon
of SSB occurs. Finally, for $\Gamma=-0.5$, the BEC is essentially
localized in the right well. For values of $\Gamma<-0.7$ the equation
does not predict a possible solution, i.e., we get a singularity from
the numerical results and the configuration of the condensate collapses
in a nonphysical state.

Further, we analyzed the behavior of the solution of the effective
NPSE \eqref{1DNPSEcalixto} for different values of $V_{0}$ and the
modulus of $\Gamma$, in order to establish a phase diagram for the
BEC states. The results are displayed in Fig. \ref{diagramanpse}.
Note that the region presenting symmetrical phase (J) shrinks as we
increase the $\varepsilon$ value, while the region with collapsed
states (C) increases. However, the region presenting asymmetric solutions
(SSB) does not significantly change its shape but presents a shift
to smaller $\Gamma$ values as $\varepsilon$ increases. This result
emphasizes the importance of controlling the transverse confinement
on the solutions to be obtained.

\begin{figure*}[tb]
\centering \includegraphics[width=0.9\textwidth]{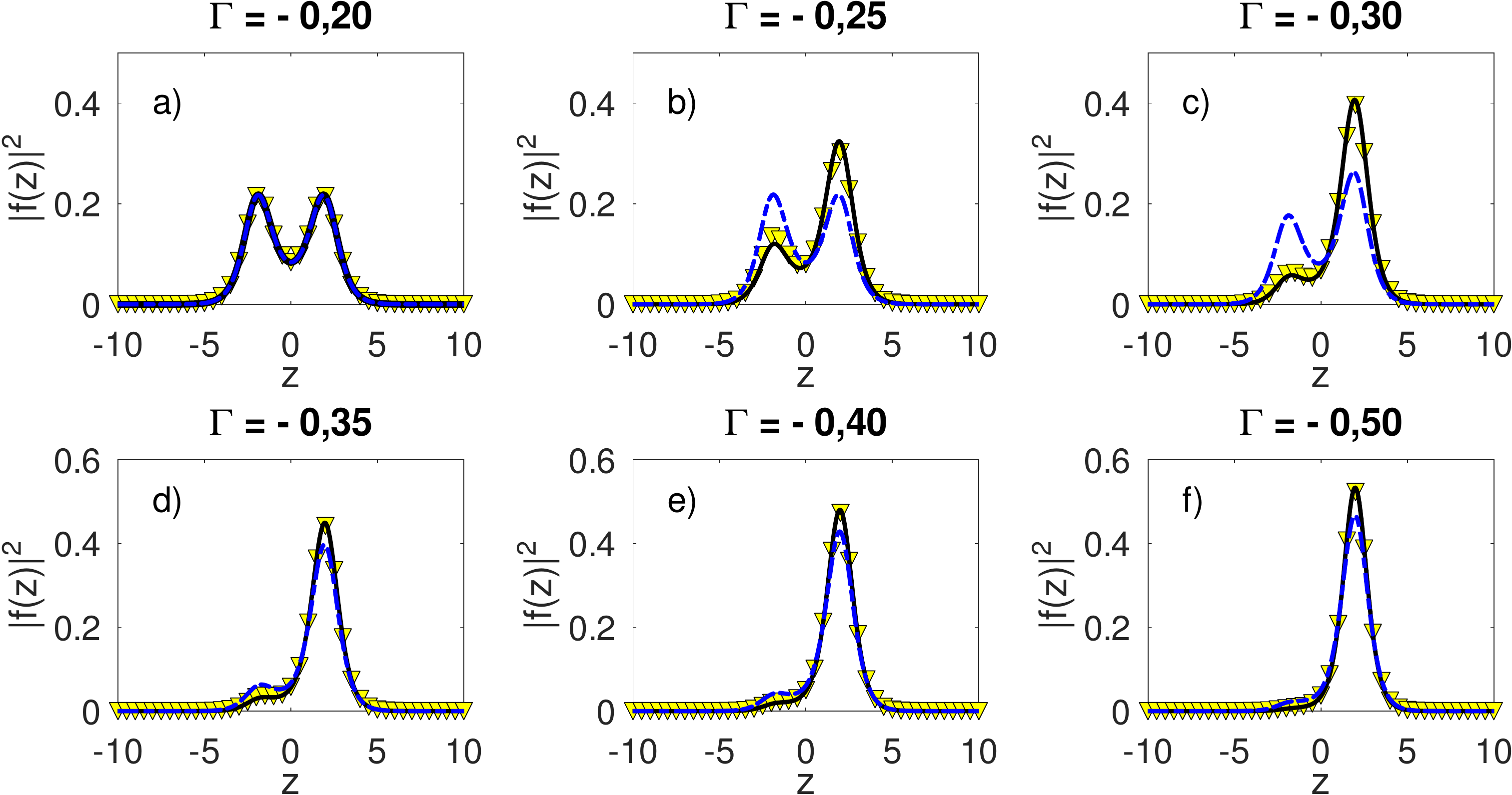} \caption{Density profile obtained from the NPSE (black solid line), cubic NLSE
(blue dashed line), and GP equation (yellow triangles) by considering
different values of interaction strengths ($\Gamma$).}
 \label{compcubicnpse3D}
\end{figure*}

Next, in order to compare our results with those from the other equations,
we investigated the solutions of GP equation and cubic NLSE, represented
respectively by Eqs. \eqref{GPE3D} and \eqref{cubicNLSE}, carrying
out the calculations for various values of $\Gamma$, keeping the
value $V_{0}=0.8\hbar\omega_{\perp}$. The results are summarized
in Fig. \ref{compcubicnpse3D}. The most relevant result is the confirmation
of the symmetry breaking by the GP equation in a perfect accuracy
with the results obtained from the effective NPSE. It is observed
that the cubic NLSE reveals an inaccuracy regarding to the GP equation
for all asymmetric cases. For example, in Fig. \ref{compcubicnpse3D}(b)
with $\Gamma=-0.25$ the cubic equation indicates an axial probability
density, $|f(z)|^{2}$, symmetric about the origin while the GP equation
clearly refers to a symmetry breaking, likewise to the effective equation,
demonstrating better accuracy than the previous one. 

The above results ensure the accuracy of the effective equation against
the results obtained for the ground states of the GP equation. Following,
we carried out calculations to attest the stability of our results
and to analyze the dynamics of the density profile of the BEC. To
this end, we use a real-time evolution method that uses as input solution
the ground-state obtained via imaginary-time propagation method plus
a $10\%$ of white noise in its amplitude profile. Next, we consider
two different situations in our analyses, namely, unaltered double-well
potential (i.e., as described by Eq. \eqref{asdf}) and a double-well
with sudden displacement in its depth ($V_{0}$).

\begin{figure*}[tb]
\centering \includegraphics[width=0.9\textwidth]{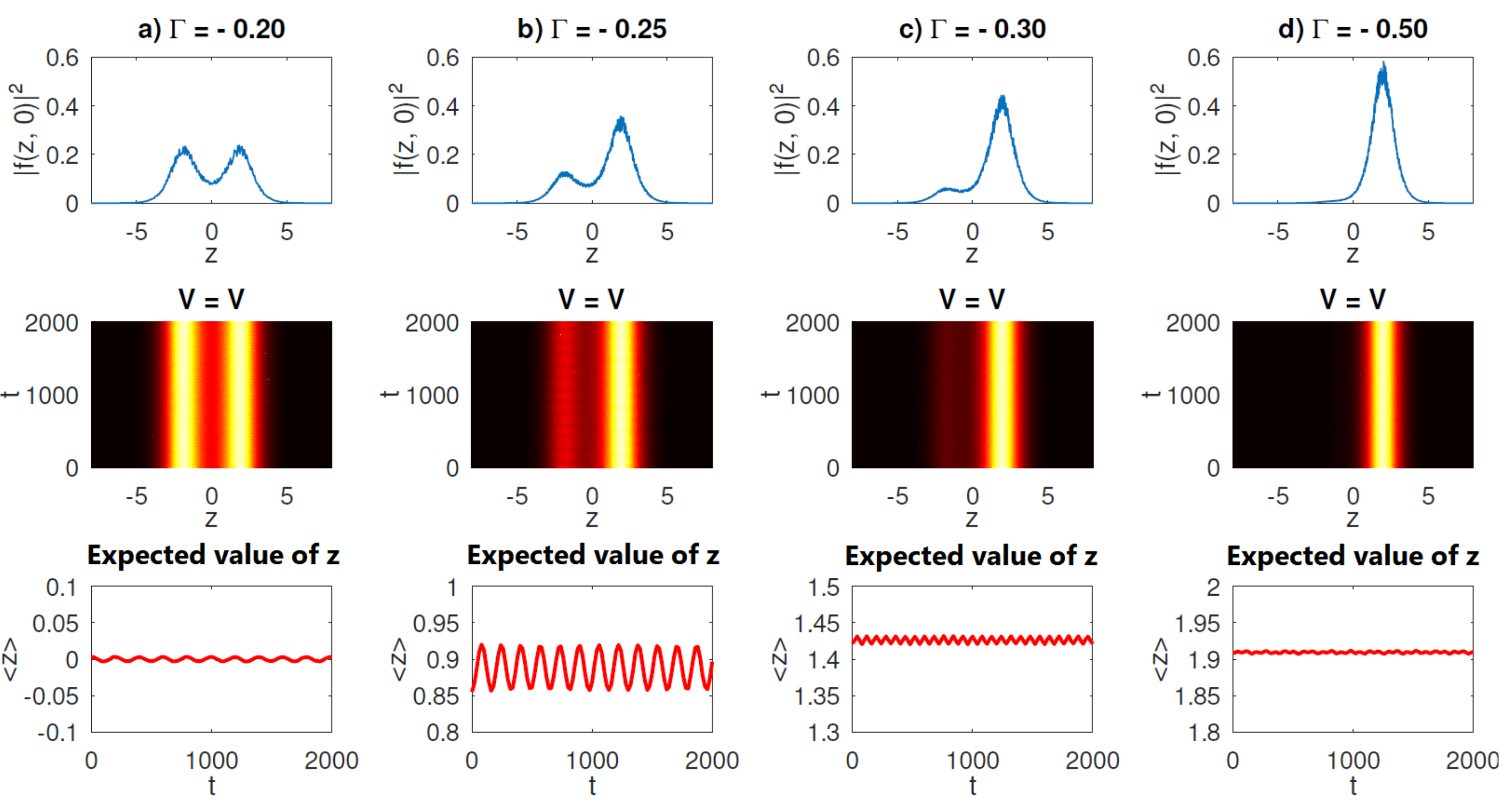} \caption{Real time simulations of perturbed ground states of NPSE for different
values of $\Gamma$ (represented by the upper panels (a)-(d)). In
the middle panels the 3D profiles of the evolution of the input solutions
are shown, while in the respective lower panels we present the behavior
of the temporal evolution of the mean value of z ($\langle z\rangle$).}
 \label{Vfixo}
\end{figure*}

The results of the case dealing with the unaltered potential are shown
in Fig. \ref{Vfixo}. In the upper panels we exhibit the input profiles,
with (a) being a symmetric input state while in (b), (c) and (d) we
have asymmetric input states. It is observed in the central panels
the stability of the propagation of the solutions up to $t=2000$
and in the lower panels the corresponding small oscillations of expected
value of z (i.e., $\langle z\rangle$). As a consequence of these
results, one can observe that all states are dynamically stable in
the face of perturbations added to the ground states analyzed here.

\begin{figure*}[tb]
\centering \includegraphics[width=0.9\textwidth]{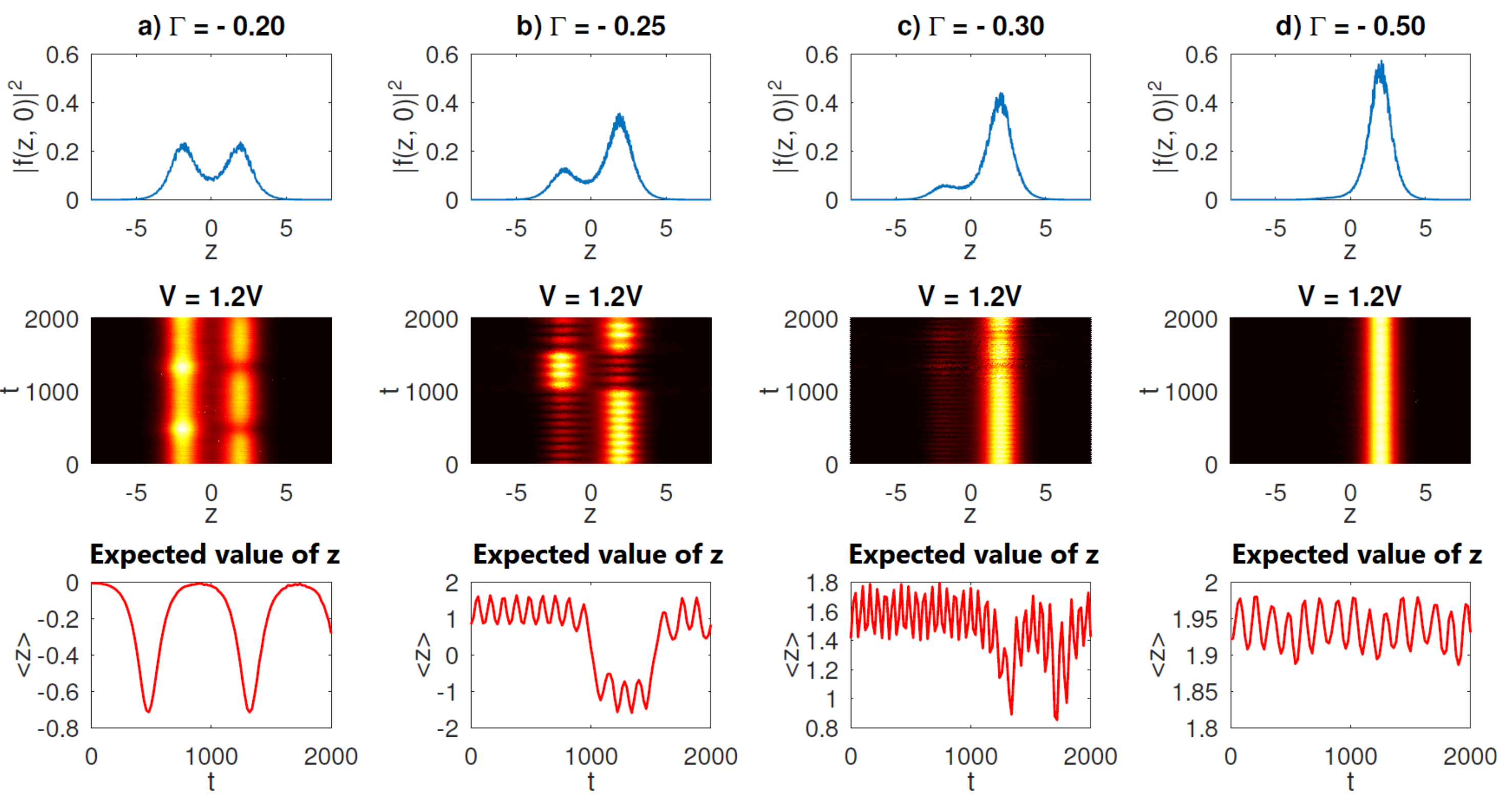} \caption{Real time simulations of perturbed ground states of NPSE with a boosted
amplitude double-well potential for different values of $\Gamma$
(represented by the upper panels (a)-(d)). In the middle panels the
3D profiles of the evolution of the input solutions are shown, while
in the respective lower panels we present the behavior of the temporal
evolution of the mean value of $z$ ($\langle z\rangle$).}
 \label{Valterado}
\end{figure*}

\qquad{}In the second case of analyzing the dynamics of the solutions,
we include an instantaneous increase in the depth of the potential,
given by:
\begin{equation}
V\rightarrow1.2\times\left(V_{\mathrm{DW}}\right).\label{12V}
\end{equation}
This sudden adjustment caused more noticeable periodic oscillations
in our system, as can be seen in the results displayed in Fig. \ref{Valterado}.
We used the same $\Gamma$ values for the input solutions and analyzed
the evolution in real time evolution under the same parameters as
in the previous case. Despite the oscillations in the expected value
of z (lower panels), it appears that the solutions are robust by observing
the evolutions obtained in the central panels. By observing the column
(a) of Fig. \ref{Valterado}, which represents the evolution of the
symmetric state, it is interesting to note that the figure in the
central panel has its maximum oscillating for negative values of $z$,
that is, going to the left well and restoring the initial position
after some time, displaying a pattern like Josephson oscillations
\cite{MAZZARELLA_JPB09,Mazzarella_PRA10}. In fact, the well in which
the system is headed is arbitrary and here we represent only the cases
in which the solution led to the right. In column (b) we have an asymmetric
input solution, with a higher probability density on the right well,
and after some time the oscillations of the particles between the
two wells can be seen, highlighted by the variation of the potential
module, when compared to the same evolution with the fixed potential.
This behavior is emphasized by the expected value of $z$, shown in
the lower panel. Finally, in columns (c) and (d) we present the evolution
of the profiles obtained with $\Gamma=-0.30$ and $-0.50$, respectively.
It is important to note that decreasing self-interaction $\Gamma$
decreases the amplitude of $\langle z\rangle$ oscillations. For example,
in (c) the maximum percentage variation of the expected value $|\langle z\rangle_{maximum}-\langle z\rangle_{minimum}|/\langle z\rangle_{maximum}$
is 54\%, while (d) presents approximately 5\%.

\section{Conclusion}

In conclusion, we investigated the influence of a transversal funnel-like
potential on the SSB inherent to the longitudinal profile of a BEC.
We derived an effective NPSE that accurately describes the longitudinal
profile of the system, for which we analyze the SSB of the wave function.
The symmetry break was observed for several interaction strength values
as a function of the minimum potential well. Our results confirmed
that this system experiences a symmetry break in perfect agreement
with the results obtained from the 3D GP equation. We observed a quantum
phase diagram presenting three distinct types of solutions, i.e.,
symmetric phase (Josephson phase), the SSB phase and the values representing
the collapsed state (unstable states). Furthermore, we analyzed the
obtained symmetric and asymmetric solutions by using a real-time evolution
method, in which it was possible to confirm the stability of the results.

These results show that the form of confinement of the BEC can change
static and dynamic patterns associated with SSB of the system. In
addition, the effective NPSE derived in Ref. \cite{Santos_JPB19}
can be used to correctly describe the system even in the presence
of SSB.
\begin{acknowledgments}
The author acknowledges the financial support of the Brazilian agencies
CNPq (\#306065/2019-3 \& \#425718/2018-2), CAPES, and FAPEG (PRONEM
\#201710267000540 \& PRONEX \#201710267000503). This work was also
performed as part of the Brazilian National Institute of Science and
Technology (INCT) for Quantum Information (\#465469/2014-0). WBC also
thanks Juracy Leandro dos Santos for his contribution in implementing
part of the infrastructure used in the simulation processes.
\end{acknowledgments}

\bibliographystyle{apsrev4-1}
\bibliography{Refs}

\begin{thebibliography}{37}%
\makeatletter
\providecommand \@ifxundefined [1]{%
 \@ifx{#1\undefined}
}%
\providecommand \@ifnum [1]{%
 \ifnum #1\expandafter \@firstoftwo
 \else \expandafter \@secondoftwo
 \fi
}%
\providecommand \@ifx [1]{%
 \ifx #1\expandafter \@firstoftwo
 \else \expandafter \@secondoftwo
 \fi
}%
\providecommand \natexlab [1]{#1}%
\providecommand \enquote  [1]{``#1''}%
\providecommand \bibnamefont  [1]{#1}%
\providecommand \bibfnamefont [1]{#1}%
\providecommand \citenamefont [1]{#1}%
\providecommand \href@noop [0]{\@secondoftwo}%
\providecommand \href [0]{\begingroup \@sanitize@url \@href}%
\providecommand \@href[1]{\@@startlink{#1}\@@href}%
\providecommand \@@href[1]{\endgroup#1\@@endlink}%
\providecommand \@sanitize@url [0]{\catcode `\\12\catcode `\$12\catcode
  `\&12\catcode `\#12\catcode `\^12\catcode `\_12\catcode `\%12\relax}%
\providecommand \@@startlink[1]{}%
\providecommand \@@endlink[0]{}%
\providecommand \url  [0]{\begingroup\@sanitize@url \@url }%
\providecommand \@url [1]{\endgroup\@href {#1}{\urlprefix }}%
\providecommand \urlprefix  [0]{URL }%
\providecommand \Eprint [0]{\href }%
\providecommand \doibase [0]{http://dx.doi.org/}%
\providecommand \selectlanguage [0]{\@gobble}%
\providecommand \bibinfo  [0]{\@secondoftwo}%
\providecommand \bibfield  [0]{\@secondoftwo}%
\providecommand \translation [1]{[#1]}%
\providecommand \BibitemOpen [0]{}%
\providecommand \bibitemStop [0]{}%
\providecommand \bibitemNoStop [0]{.\EOS\space}%
\providecommand \EOS [0]{\spacefactor3000\relax}%
\providecommand \BibitemShut  [1]{\csname bibitem#1\endcsname}%
\let\auto@bib@innerbib\@empty
\bibitem [{\citenamefont {Anderson}\ \emph {et~al.}(1995)\citenamefont
  {Anderson}, \citenamefont {Ensher}, \citenamefont {Matthews}, \citenamefont
  {Wieman},\ and\ \citenamefont {Cornell}}]{ANDERSON_SCIENCE95}%
  \BibitemOpen
  \bibfield  {author} {\bibinfo {author} {\bibfnamefont {M.~H.}\ \bibnamefont
  {Anderson}}, \bibinfo {author} {\bibfnamefont {J.~R.}\ \bibnamefont
  {Ensher}}, \bibinfo {author} {\bibfnamefont {M.~R.}\ \bibnamefont
  {Matthews}}, \bibinfo {author} {\bibfnamefont {C.~E.}\ \bibnamefont
  {Wieman}}, \ and\ \bibinfo {author} {\bibfnamefont {E.~A.}\ \bibnamefont
  {Cornell}},\ }\href {\doibase 10.1126/science.269.5221.198} {\bibfield
  {journal} {\bibinfo  {journal} {Science}\ }\textbf {\bibinfo {volume}
  {269}},\ \bibinfo {pages} {198} (\bibinfo {year} {1995})}\BibitemShut
  {NoStop}%
\bibitem [{\citenamefont {Davis}\ \emph {et~al.}(1995)\citenamefont {Davis},
  \citenamefont {Mewes}, \citenamefont {Andrews}, \citenamefont {van Druten},
  \citenamefont {Durfee}, \citenamefont {Kurn},\ and\ \citenamefont
  {Ketterle}}]{DAVIS_PRL95}%
  \BibitemOpen
  \bibfield  {author} {\bibinfo {author} {\bibfnamefont {K.~B.}\ \bibnamefont
  {Davis}}, \bibinfo {author} {\bibfnamefont {M.~O.}\ \bibnamefont {Mewes}},
  \bibinfo {author} {\bibfnamefont {M.~R.}\ \bibnamefont {Andrews}}, \bibinfo
  {author} {\bibfnamefont {N.~J.}\ \bibnamefont {van Druten}}, \bibinfo
  {author} {\bibfnamefont {D.~S.}\ \bibnamefont {Durfee}}, \bibinfo {author}
  {\bibfnamefont {D.~M.}\ \bibnamefont {Kurn}}, \ and\ \bibinfo {author}
  {\bibfnamefont {W.}~\bibnamefont {Ketterle}},\ }\href {\doibase
  10.1103/PhysRevLett.75.3969} {\bibfield  {journal} {\bibinfo  {journal}
  {Phys. Rev. Lett.}\ }\textbf {\bibinfo {volume} {75}},\ \bibinfo {pages}
  {3969} (\bibinfo {year} {1995})}\BibitemShut {NoStop}%
\bibitem [{\citenamefont {Bradley}\ \emph {et~al.}(1995)\citenamefont
  {Bradley}, \citenamefont {Sackett}, \citenamefont {Tollett},\ and\
  \citenamefont {Hulet}}]{BRADLEY_PRL95}%
  \BibitemOpen
  \bibfield  {author} {\bibinfo {author} {\bibfnamefont {C.~C.}\ \bibnamefont
  {Bradley}}, \bibinfo {author} {\bibfnamefont {C.~A.}\ \bibnamefont
  {Sackett}}, \bibinfo {author} {\bibfnamefont {J.~J.}\ \bibnamefont
  {Tollett}}, \ and\ \bibinfo {author} {\bibfnamefont {R.~G.}\ \bibnamefont
  {Hulet}},\ }\href {\doibase 10.1103/PhysRevLett.75.1687} {\bibfield
  {journal} {\bibinfo  {journal} {Phys. Rev. Lett.}\ }\textbf {\bibinfo
  {volume} {75}},\ \bibinfo {pages} {1687} (\bibinfo {year}
  {1995})}\BibitemShut {NoStop}%
\bibitem [{\citenamefont {Burger}\ \emph {et~al.}(1999)\citenamefont {Burger},
  \citenamefont {Bongs}, \citenamefont {Dettmer}, \citenamefont {Ertmer},
  \citenamefont {Sengstock}, \citenamefont {Sanpera}, \citenamefont
  {Shlyapnikov},\ and\ \citenamefont {Lewenstein}}]{BURGER_PRL99}%
  \BibitemOpen
  \bibfield  {author} {\bibinfo {author} {\bibfnamefont {S.}~\bibnamefont
  {Burger}}, \bibinfo {author} {\bibfnamefont {K.}~\bibnamefont {Bongs}},
  \bibinfo {author} {\bibfnamefont {S.}~\bibnamefont {Dettmer}}, \bibinfo
  {author} {\bibfnamefont {W.}~\bibnamefont {Ertmer}}, \bibinfo {author}
  {\bibfnamefont {K.}~\bibnamefont {Sengstock}}, \bibinfo {author}
  {\bibfnamefont {A.}~\bibnamefont {Sanpera}}, \bibinfo {author} {\bibfnamefont
  {G.~V.}\ \bibnamefont {Shlyapnikov}}, \ and\ \bibinfo {author} {\bibfnamefont
  {M.}~\bibnamefont {Lewenstein}},\ }\href {\doibase
  10.1103/PhysRevLett.83.5198} {\bibfield  {journal} {\bibinfo  {journal}
  {Phys. Rev. Lett.}\ }\textbf {\bibinfo {volume} {83}},\ \bibinfo {pages}
  {5198} (\bibinfo {year} {1999})}\BibitemShut {NoStop}%
\bibitem [{\citenamefont {{Kh. Abdullaev}}\ \emph {et~al.}(2005)\citenamefont
  {{Kh. Abdullaev}}, \citenamefont {Gammal}, \citenamefont {Kamchatnov},\ and\
  \citenamefont {Tomio}}]{ABDULLAEV_IJMPB05}%
  \BibitemOpen
  \bibfield  {author} {\bibinfo {author} {\bibfnamefont {F.}~\bibnamefont {{Kh.
  Abdullaev}}}, \bibinfo {author} {\bibfnamefont {A.}~\bibnamefont {Gammal}},
  \bibinfo {author} {\bibfnamefont {A.~M.}\ \bibnamefont {Kamchatnov}}, \ and\
  \bibinfo {author} {\bibfnamefont {L.}~\bibnamefont {Tomio}},\ }\href
  {\doibase 10.1142/S0217979205032279} {\bibfield  {journal} {\bibinfo
  {journal} {Int. J. Mod. Phys. B}\ }\textbf {\bibinfo {volume} {19}},\
  \bibinfo {pages} {3415} (\bibinfo {year} {2005})}\BibitemShut {NoStop}%
\bibitem [{\citenamefont {Cornish}\ \emph {et~al.}(2006)\citenamefont
  {Cornish}, \citenamefont {Thompson},\ and\ \citenamefont
  {Wieman}}]{CORNISH_PRL06}%
  \BibitemOpen
  \bibfield  {author} {\bibinfo {author} {\bibfnamefont {S.~L.}\ \bibnamefont
  {Cornish}}, \bibinfo {author} {\bibfnamefont {S.~T.}\ \bibnamefont
  {Thompson}}, \ and\ \bibinfo {author} {\bibfnamefont {C.~E.}\ \bibnamefont
  {Wieman}},\ }\href {\doibase 10.1103/PhysRevLett.96.170401} {\bibfield
  {journal} {\bibinfo  {journal} {Phys. Rev. Lett.}\ }\textbf {\bibinfo
  {volume} {96}},\ \bibinfo {pages} {170401} (\bibinfo {year}
  {2006})}\BibitemShut {NoStop}%
\bibitem [{\citenamefont {Khaykovich}\ \emph {et~al.}(2002)\citenamefont
  {Khaykovich}, \citenamefont {Schreck}, \citenamefont {Ferrari}, \citenamefont
  {Bourdel}, \citenamefont {Cubizolles}, \citenamefont {Carr}, \citenamefont
  {Castin},\ and\ \citenamefont {Salomon}}]{KHAYKOVICH_SCIENCE02}%
  \BibitemOpen
  \bibfield  {author} {\bibinfo {author} {\bibfnamefont {L.}~\bibnamefont
  {Khaykovich}}, \bibinfo {author} {\bibfnamefont {F.}~\bibnamefont {Schreck}},
  \bibinfo {author} {\bibfnamefont {G.}~\bibnamefont {Ferrari}}, \bibinfo
  {author} {\bibfnamefont {T.}~\bibnamefont {Bourdel}}, \bibinfo {author}
  {\bibfnamefont {J.}~\bibnamefont {Cubizolles}}, \bibinfo {author}
  {\bibfnamefont {L.~D.}\ \bibnamefont {Carr}}, \bibinfo {author}
  {\bibfnamefont {Y.}~\bibnamefont {Castin}}, \ and\ \bibinfo {author}
  {\bibfnamefont {C.}~\bibnamefont {Salomon}},\ }\href {\doibase
  10.1126/science.1071021} {\bibfield  {journal} {\bibinfo  {journal}
  {Science}\ }\textbf {\bibinfo {volume} {296}},\ \bibinfo {pages} {1290}
  (\bibinfo {year} {2002})}\BibitemShut {NoStop}%
\bibitem [{\citenamefont {Matthews}\ \emph {et~al.}(1999)\citenamefont
  {Matthews}, \citenamefont {Anderson}, \citenamefont {Haljan}, \citenamefont
  {Hall}, \citenamefont {Wieman},\ and\ \citenamefont
  {Cornell}}]{MATTHEWS_PRL99}%
  \BibitemOpen
  \bibfield  {author} {\bibinfo {author} {\bibfnamefont {M.~R.}\ \bibnamefont
  {Matthews}}, \bibinfo {author} {\bibfnamefont {B.~P.}\ \bibnamefont
  {Anderson}}, \bibinfo {author} {\bibfnamefont {P.~C.}\ \bibnamefont
  {Haljan}}, \bibinfo {author} {\bibfnamefont {D.~S.}\ \bibnamefont {Hall}},
  \bibinfo {author} {\bibfnamefont {C.~E.}\ \bibnamefont {Wieman}}, \ and\
  \bibinfo {author} {\bibfnamefont {E.~A.}\ \bibnamefont {Cornell}},\ }\href
  {\doibase 10.1103/PhysRevLett.83.2498} {\bibfield  {journal} {\bibinfo
  {journal} {Phys. Rev. Lett.}\ }\textbf {\bibinfo {volume} {83}},\ \bibinfo
  {pages} {2498} (\bibinfo {year} {1999})}\BibitemShut {NoStop}%
\bibitem [{\citenamefont {Madison}\ \emph {et~al.}(2000)\citenamefont
  {Madison}, \citenamefont {Chevy}, \citenamefont {Wohlleben},\ and\
  \citenamefont {Dalibard}}]{MADISON_PRL00}%
  \BibitemOpen
  \bibfield  {author} {\bibinfo {author} {\bibfnamefont {K.~W.}\ \bibnamefont
  {Madison}}, \bibinfo {author} {\bibfnamefont {F.}~\bibnamefont {Chevy}},
  \bibinfo {author} {\bibfnamefont {W.}~\bibnamefont {Wohlleben}}, \ and\
  \bibinfo {author} {\bibfnamefont {J.}~\bibnamefont {Dalibard}},\ }\href
  {\doibase 10.1103/PhysRevLett.84.806} {\bibfield  {journal} {\bibinfo
  {journal} {Phys. Rev. Lett.}\ }\textbf {\bibinfo {volume} {84}},\ \bibinfo
  {pages} {806} (\bibinfo {year} {2000})}\BibitemShut {NoStop}%
\bibitem [{\citenamefont {Lin}\ \emph {et~al.}(2011)\citenamefont {Lin},
  \citenamefont {Jim{\'{e}}nez-Garc{\'{i}}a},\ and\ \citenamefont
  {Spielman}}]{Lin_NATURE11}%
  \BibitemOpen
  \bibfield  {author} {\bibinfo {author} {\bibfnamefont {Y.-J.}\ \bibnamefont
  {Lin}}, \bibinfo {author} {\bibfnamefont {K.}~\bibnamefont
  {Jim{\'{e}}nez-Garc{\'{i}}a}}, \ and\ \bibinfo {author} {\bibfnamefont
  {I.~B.}\ \bibnamefont {Spielman}},\ }\href {\doibase 10.1038/nature09887}
  {\bibfield  {journal} {\bibinfo  {journal} {Nature}\ }\textbf {\bibinfo
  {volume} {471}},\ \bibinfo {pages} {83} (\bibinfo {year} {2011})}\BibitemShut
  {NoStop}%
\bibitem [{\citenamefont {Billy}\ \emph {et~al.}(2008)\citenamefont {Billy},
  \citenamefont {Josse}, \citenamefont {Zuo}, \citenamefont {Bernard},
  \citenamefont {Hambrecht}, \citenamefont {Lugan}, \citenamefont
  {Cl{\'{e}}ment}, \citenamefont {Sanchez-Palencia}, \citenamefont {Bouyer},\
  and\ \citenamefont {Aspect}}]{BILLY_NATURE08}%
  \BibitemOpen
  \bibfield  {author} {\bibinfo {author} {\bibfnamefont {J.}~\bibnamefont
  {Billy}}, \bibinfo {author} {\bibfnamefont {V.}~\bibnamefont {Josse}},
  \bibinfo {author} {\bibfnamefont {Z.}~\bibnamefont {Zuo}}, \bibinfo {author}
  {\bibfnamefont {A.}~\bibnamefont {Bernard}}, \bibinfo {author} {\bibfnamefont
  {B.}~\bibnamefont {Hambrecht}}, \bibinfo {author} {\bibfnamefont
  {P.}~\bibnamefont {Lugan}}, \bibinfo {author} {\bibfnamefont
  {D.}~\bibnamefont {Cl{\'{e}}ment}}, \bibinfo {author} {\bibfnamefont
  {L.}~\bibnamefont {Sanchez-Palencia}}, \bibinfo {author} {\bibfnamefont
  {P.}~\bibnamefont {Bouyer}}, \ and\ \bibinfo {author} {\bibfnamefont
  {A.}~\bibnamefont {Aspect}},\ }\href {\doibase 10.1038/nature07000}
  {\bibfield  {journal} {\bibinfo  {journal} {Nature}\ }\textbf {\bibinfo
  {volume} {453}},\ \bibinfo {pages} {891} (\bibinfo {year}
  {2008})}\BibitemShut {NoStop}%
\bibitem [{\citenamefont {Roati}\ \emph {et~al.}(2008)\citenamefont {Roati},
  \citenamefont {D'Errico}, \citenamefont {Fallani}, \citenamefont {Fattori},
  \citenamefont {Fort}, \citenamefont {Zaccanti}, \citenamefont {Modugno},
  \citenamefont {Modugno},\ and\ \citenamefont {Inguscio}}]{ROATI_NATURE08}%
  \BibitemOpen
  \bibfield  {author} {\bibinfo {author} {\bibfnamefont {G.}~\bibnamefont
  {Roati}}, \bibinfo {author} {\bibfnamefont {C.}~\bibnamefont {D'Errico}},
  \bibinfo {author} {\bibfnamefont {L.}~\bibnamefont {Fallani}}, \bibinfo
  {author} {\bibfnamefont {M.}~\bibnamefont {Fattori}}, \bibinfo {author}
  {\bibfnamefont {C.}~\bibnamefont {Fort}}, \bibinfo {author} {\bibfnamefont
  {M.}~\bibnamefont {Zaccanti}}, \bibinfo {author} {\bibfnamefont
  {G.}~\bibnamefont {Modugno}}, \bibinfo {author} {\bibfnamefont
  {M.}~\bibnamefont {Modugno}}, \ and\ \bibinfo {author} {\bibfnamefont
  {M.}~\bibnamefont {Inguscio}},\ }\href {\doibase 10.1038/nature07071}
  {\bibfield  {journal} {\bibinfo  {journal} {Nature}\ }\textbf {\bibinfo
  {volume} {453}},\ \bibinfo {pages} {895} (\bibinfo {year}
  {2008})}\BibitemShut {NoStop}%
\bibitem [{\citenamefont {Cabrera}\ \emph {et~al.}(2018)\citenamefont
  {Cabrera}, \citenamefont {Tanzi}, \citenamefont {Sanz}, \citenamefont
  {Naylor}, \citenamefont {Thomas}, \citenamefont {Cheiney},\ and\
  \citenamefont {Tarruell}}]{CABRERA_SCIENCE18}%
  \BibitemOpen
  \bibfield  {author} {\bibinfo {author} {\bibfnamefont {C.~R.}\ \bibnamefont
  {Cabrera}}, \bibinfo {author} {\bibfnamefont {L.}~\bibnamefont {Tanzi}},
  \bibinfo {author} {\bibfnamefont {J.}~\bibnamefont {Sanz}}, \bibinfo {author}
  {\bibfnamefont {B.}~\bibnamefont {Naylor}}, \bibinfo {author} {\bibfnamefont
  {P.}~\bibnamefont {Thomas}}, \bibinfo {author} {\bibfnamefont
  {P.}~\bibnamefont {Cheiney}}, \ and\ \bibinfo {author} {\bibfnamefont
  {L.}~\bibnamefont {Tarruell}},\ }\href {\doibase 10.1126/science.aao5686}
  {\bibfield  {journal} {\bibinfo  {journal} {Science}\ }\textbf {\bibinfo
  {volume} {359}},\ \bibinfo {pages} {301} (\bibinfo {year}
  {2018})}\BibitemShut {NoStop}%
\bibitem [{\citenamefont {Cheiney}\ \emph {et~al.}(2018)\citenamefont
  {Cheiney}, \citenamefont {Cabrera}, \citenamefont {Sanz}, \citenamefont
  {Naylor}, \citenamefont {Tanzi},\ and\ \citenamefont
  {Tarruell}}]{CHEINEY_PRL18}%
  \BibitemOpen
  \bibfield  {author} {\bibinfo {author} {\bibfnamefont {P.}~\bibnamefont
  {Cheiney}}, \bibinfo {author} {\bibfnamefont {C.~R.}\ \bibnamefont
  {Cabrera}}, \bibinfo {author} {\bibfnamefont {J.}~\bibnamefont {Sanz}},
  \bibinfo {author} {\bibfnamefont {B.}~\bibnamefont {Naylor}}, \bibinfo
  {author} {\bibfnamefont {L.}~\bibnamefont {Tanzi}}, \ and\ \bibinfo {author}
  {\bibfnamefont {L.}~\bibnamefont {Tarruell}},\ }\href {\doibase
  10.1103/PhysRevLett.120.135301} {\bibfield  {journal} {\bibinfo  {journal}
  {Phys. Rev. Lett.}\ }\textbf {\bibinfo {volume} {120}},\ \bibinfo {pages}
  {135301} (\bibinfo {year} {2018})}\BibitemShut {NoStop}%
\bibitem [{\citenamefont {Semeghini}\ \emph {et~al.}(2018)\citenamefont
  {Semeghini}, \citenamefont {Ferioli}, \citenamefont {Masi}, \citenamefont
  {Mazzinghi}, \citenamefont {Wolswijk}, \citenamefont {Minardi}, \citenamefont
  {Modugno}, \citenamefont {Modugno}, \citenamefont {Inguscio},\ and\
  \citenamefont {Fattori}}]{SEMEGHINI_PRL18}%
  \BibitemOpen
  \bibfield  {author} {\bibinfo {author} {\bibfnamefont {G.}~\bibnamefont
  {Semeghini}}, \bibinfo {author} {\bibfnamefont {G.}~\bibnamefont {Ferioli}},
  \bibinfo {author} {\bibfnamefont {L.}~\bibnamefont {Masi}}, \bibinfo {author}
  {\bibfnamefont {C.}~\bibnamefont {Mazzinghi}}, \bibinfo {author}
  {\bibfnamefont {L.}~\bibnamefont {Wolswijk}}, \bibinfo {author}
  {\bibfnamefont {F.}~\bibnamefont {Minardi}}, \bibinfo {author} {\bibfnamefont
  {M.}~\bibnamefont {Modugno}}, \bibinfo {author} {\bibfnamefont
  {G.}~\bibnamefont {Modugno}}, \bibinfo {author} {\bibfnamefont
  {M.}~\bibnamefont {Inguscio}}, \ and\ \bibinfo {author} {\bibfnamefont
  {M.}~\bibnamefont {Fattori}},\ }\href {\doibase
  10.1103/PhysRevLett.120.235301} {\bibfield  {journal} {\bibinfo  {journal}
  {Phys. Rev. Lett.}\ }\textbf {\bibinfo {volume} {120}},\ \bibinfo {pages}
  {235301} (\bibinfo {year} {2018})}\BibitemShut {NoStop}%
\bibitem [{\citenamefont {D'Errico}\ \emph {et~al.}(2019)\citenamefont
  {D'Errico}, \citenamefont {Burchianti}, \citenamefont {Prevedelli},
  \citenamefont {Salasnich}, \citenamefont {Ancilotto}, \citenamefont
  {Modugno}, \citenamefont {Minardi},\ and\ \citenamefont
  {Fort}}]{DERRICO_PRR19}%
  \BibitemOpen
  \bibfield  {author} {\bibinfo {author} {\bibfnamefont {C.}~\bibnamefont
  {D'Errico}}, \bibinfo {author} {\bibfnamefont {A.}~\bibnamefont
  {Burchianti}}, \bibinfo {author} {\bibfnamefont {M.}~\bibnamefont
  {Prevedelli}}, \bibinfo {author} {\bibfnamefont {L.}~\bibnamefont
  {Salasnich}}, \bibinfo {author} {\bibfnamefont {F.}~\bibnamefont
  {Ancilotto}}, \bibinfo {author} {\bibfnamefont {M.}~\bibnamefont {Modugno}},
  \bibinfo {author} {\bibfnamefont {F.}~\bibnamefont {Minardi}}, \ and\
  \bibinfo {author} {\bibfnamefont {C.}~\bibnamefont {Fort}},\ }\href {\doibase
  10.1103/PhysRevResearch.1.033155} {\bibfield  {journal} {\bibinfo  {journal}
  {Phys. Rev. Res.}\ }\textbf {\bibinfo {volume} {1}},\ \bibinfo {pages}
  {033155} (\bibinfo {year} {2019})}\BibitemShut {NoStop}%
\bibitem [{\citenamefont {Yuan}\ and\ \citenamefont {Ding}(2005)}]{Yuan_PLA05}%
  \BibitemOpen
  \bibfield  {author} {\bibinfo {author} {\bibfnamefont {Q.-X.}\ \bibnamefont
  {Yuan}}\ and\ \bibinfo {author} {\bibfnamefont {G.-H.}\ \bibnamefont
  {Ding}},\ }\href {\doibase 10.1016/j.physleta.2005.06.057} {\bibfield
  {journal} {\bibinfo  {journal} {Phys. Lett. A}\ }\textbf {\bibinfo {volume}
  {344}},\ \bibinfo {pages} {156} (\bibinfo {year} {2005})}\BibitemShut
  {NoStop}%
\bibitem [{\citenamefont {Xie}\ \emph {et~al.}(2015)\citenamefont {Xie},
  \citenamefont {Liu},\ and\ \citenamefont {Rong}}]{Xie_MPLB15}%
  \BibitemOpen
  \bibfield  {author} {\bibinfo {author} {\bibfnamefont {Q.}~\bibnamefont
  {Xie}}, \bibinfo {author} {\bibfnamefont {X.}~\bibnamefont {Liu}}, \ and\
  \bibinfo {author} {\bibfnamefont {S.}~\bibnamefont {Rong}},\ }\href {\doibase
  10.1142/S021798491550150X} {\bibfield  {journal} {\bibinfo  {journal} {Mod.
  Phys. Lett. B}\ }\textbf {\bibinfo {volume} {29}},\ \bibinfo {pages}
  {1550150} (\bibinfo {year} {2015})}\BibitemShut {NoStop}%
\bibitem [{\citenamefont {Shchesnovich}\ \emph {et~al.}(2004)\citenamefont
  {Shchesnovich}, \citenamefont {Malomed},\ and\ \citenamefont
  {Kraenkel}}]{Shchesnovich_PD04}%
  \BibitemOpen
  \bibfield  {author} {\bibinfo {author} {\bibfnamefont {V.~S.}\ \bibnamefont
  {Shchesnovich}}, \bibinfo {author} {\bibfnamefont {B.~A.}\ \bibnamefont
  {Malomed}}, \ and\ \bibinfo {author} {\bibfnamefont {R.~A.}\ \bibnamefont
  {Kraenkel}},\ }\href {\doibase 10.1016/j.physd.2003.07.010} {\bibfield
  {journal} {\bibinfo  {journal} {Phys. D Nonlinear Phenom.}\ }\textbf
  {\bibinfo {volume} {188}},\ \bibinfo {pages} {213} (\bibinfo {year}
  {2004})}\BibitemShut {NoStop}%
\bibitem [{\citenamefont {Salasnich}\ and\ \citenamefont
  {Malomed}(2011)}]{Salasnich_MP05}%
  \BibitemOpen
  \bibfield  {author} {\bibinfo {author} {\bibfnamefont {L.}~\bibnamefont
  {Salasnich}}\ and\ \bibinfo {author} {\bibfnamefont {B.~A.}\ \bibnamefont
  {Malomed}},\ }\href {\doibase 10.1080/00268976.2011.602370} {\bibfield
  {journal} {\bibinfo  {journal} {Mol. Phys.}\ }\textbf {\bibinfo {volume}
  {109}},\ \bibinfo {pages} {2737} (\bibinfo {year} {2011})}\BibitemShut
  {NoStop}%
\bibitem [{\citenamefont {Mazzarella}\ \emph {et~al.}(2009)\citenamefont
  {Mazzarella}, \citenamefont {Moratti}, \citenamefont {Salasnich},
  \citenamefont {Salerno},\ and\ \citenamefont {Toigo}}]{MAZZARELLA_JPB09}%
  \BibitemOpen
  \bibfield  {author} {\bibinfo {author} {\bibfnamefont {G.}~\bibnamefont
  {Mazzarella}}, \bibinfo {author} {\bibfnamefont {M.}~\bibnamefont {Moratti}},
  \bibinfo {author} {\bibfnamefont {L.}~\bibnamefont {Salasnich}}, \bibinfo
  {author} {\bibfnamefont {M.}~\bibnamefont {Salerno}}, \ and\ \bibinfo
  {author} {\bibfnamefont {F.}~\bibnamefont {Toigo}},\ }\href {\doibase
  10.1088/0953-4075/42/12/125301} {\bibfield  {journal} {\bibinfo  {journal}
  {J. Phys. B At. Mol. Opt. Phys.}\ }\textbf {\bibinfo {volume} {42}},\
  \bibinfo {pages} {125301} (\bibinfo {year} {2009})}\BibitemShut {NoStop}%
\bibitem [{\citenamefont {Mazzarella}\ and\ \citenamefont
  {Salasnich}(2010)}]{Mazzarella_PRA10}%
  \BibitemOpen
  \bibfield  {author} {\bibinfo {author} {\bibfnamefont {G.}~\bibnamefont
  {Mazzarella}}\ and\ \bibinfo {author} {\bibfnamefont {L.}~\bibnamefont
  {Salasnich}},\ }\href {\doibase 10.1103/PhysRevA.82.033611} {\bibfield
  {journal} {\bibinfo  {journal} {Phys. Rev. A}\ }\textbf {\bibinfo {volume}
  {82}},\ \bibinfo {pages} {033611} (\bibinfo {year} {2010})}\BibitemShut
  {NoStop}%
\bibitem [{\citenamefont {Matuszewski}\ \emph {et~al.}(2007)\citenamefont
  {Matuszewski}, \citenamefont {Malomed},\ and\ \citenamefont
  {Trippenbach}}]{Matuszewski_PRA07}%
  \BibitemOpen
  \bibfield  {author} {\bibinfo {author} {\bibfnamefont {M.}~\bibnamefont
  {Matuszewski}}, \bibinfo {author} {\bibfnamefont {B.~A.}\ \bibnamefont
  {Malomed}}, \ and\ \bibinfo {author} {\bibfnamefont {M.}~\bibnamefont
  {Trippenbach}},\ }\href {\doibase 10.1103/PhysRevA.75.063621} {\bibfield
  {journal} {\bibinfo  {journal} {Phys. Rev. A}\ }\textbf {\bibinfo {volume}
  {75}},\ \bibinfo {pages} {63621} (\bibinfo {year} {2007})}\BibitemShut
  {NoStop}%
\bibitem [{\citenamefont {Mayteevarunyoo}\ \emph {et~al.}(2008)\citenamefont
  {Mayteevarunyoo}, \citenamefont {Malomed},\ and\ \citenamefont
  {Dong}}]{Mayteevarunyoo_PRA08}%
  \BibitemOpen
  \bibfield  {author} {\bibinfo {author} {\bibfnamefont {T.}~\bibnamefont
  {Mayteevarunyoo}}, \bibinfo {author} {\bibfnamefont {B.~A.}\ \bibnamefont
  {Malomed}}, \ and\ \bibinfo {author} {\bibfnamefont {G.}~\bibnamefont
  {Dong}},\ }\href {\doibase 10.1103/PhysRevA.78.053601} {\bibfield  {journal}
  {\bibinfo  {journal} {Phys. Rev. A}\ }\textbf {\bibinfo {volume} {78}},\
  \bibinfo {pages} {053601} (\bibinfo {year} {2008})}\BibitemShut {NoStop}%
\bibitem [{\citenamefont {Salasnich}\ \emph {et~al.}(2002)\citenamefont
  {Salasnich}, \citenamefont {Parola},\ and\ \citenamefont
  {Reatto}}]{SALASNICH_PRA02}%
  \BibitemOpen
  \bibfield  {author} {\bibinfo {author} {\bibfnamefont {L.}~\bibnamefont
  {Salasnich}}, \bibinfo {author} {\bibfnamefont {A.}~\bibnamefont {Parola}}, \
  and\ \bibinfo {author} {\bibfnamefont {L.}~\bibnamefont {Reatto}},\ }\href
  {\doibase 10.1103/PhysRevA.65.043614} {\bibfield  {journal} {\bibinfo
  {journal} {Phys. Rev. A}\ }\textbf {\bibinfo {volume} {65}},\ \bibinfo
  {pages} {043614} (\bibinfo {year} {2002})}\BibitemShut {NoStop}%
\bibitem [{\citenamefont {Salasnich}\ and\ \citenamefont
  {Malomed}(2009)}]{Salasnich_PRA09}%
  \BibitemOpen
  \bibfield  {author} {\bibinfo {author} {\bibfnamefont {L.}~\bibnamefont
  {Salasnich}}\ and\ \bibinfo {author} {\bibfnamefont {B.~A.}\ \bibnamefont
  {Malomed}},\ }\href {\doibase 10.1103/PhysRevA.79.053620} {\bibfield
  {journal} {\bibinfo  {journal} {Phys. Rev. A}\ }\textbf {\bibinfo {volume}
  {79}},\ \bibinfo {pages} {53620} (\bibinfo {year} {2009})}\BibitemShut
  {NoStop}%
\bibitem [{\citenamefont {Salasnich}(2009)}]{Salasnich_JPA09}%
  \BibitemOpen
  \bibfield  {author} {\bibinfo {author} {\bibfnamefont {L.}~\bibnamefont
  {Salasnich}},\ }\href {\doibase 10.1088/1751-8113/42/33/335205} {\bibfield
  {journal} {\bibinfo  {journal} {J. Phys. A Math. Theor.}\ }\textbf {\bibinfo
  {volume} {42}},\ \bibinfo {pages} {335205} (\bibinfo {year}
  {2009})}\BibitemShut {NoStop}%
\bibitem [{\citenamefont {Young-S.}\ \emph {et~al.}(2010)\citenamefont
  {Young-S.}, \citenamefont {Salasnich},\ and\ \citenamefont
  {Adhikari}}]{Young_PRA10}%
  \BibitemOpen
  \bibfield  {author} {\bibinfo {author} {\bibfnamefont {L.~E.}\ \bibnamefont
  {Young-S.}}, \bibinfo {author} {\bibfnamefont {L.}~\bibnamefont {Salasnich}},
  \ and\ \bibinfo {author} {\bibfnamefont {S.~K.}\ \bibnamefont {Adhikari}},\
  }\href {\doibase 10.1103/PhysRevA.82.053601} {\bibfield  {journal} {\bibinfo
  {journal} {Phys. Rev. A}\ }\textbf {\bibinfo {volume} {82}},\ \bibinfo
  {pages} {053601} (\bibinfo {year} {2010})}\BibitemShut {NoStop}%
\bibitem [{\citenamefont {Cardoso}\ \emph {et~al.}(2011)\citenamefont
  {Cardoso}, \citenamefont {Avelar},\ and\ \citenamefont
  {Bazeia}}]{Cardoso_PRE11}%
  \BibitemOpen
  \bibfield  {author} {\bibinfo {author} {\bibfnamefont {W.~B.~B.}\
  \bibnamefont {Cardoso}}, \bibinfo {author} {\bibfnamefont {A.~T.~T.}\
  \bibnamefont {Avelar}}, \ and\ \bibinfo {author} {\bibfnamefont
  {D.}~\bibnamefont {Bazeia}},\ }\href {\doibase 10.1103/PhysRevE.83.036604}
  {\bibfield  {journal} {\bibinfo  {journal} {Phys. Rev. E}\ }\textbf {\bibinfo
  {volume} {83}},\ \bibinfo {pages} {36604} (\bibinfo {year}
  {2011})}\BibitemShut {NoStop}%
\bibitem [{\citenamefont {Salasnich}\ and\ \citenamefont
  {Malomed}(2012)}]{Salasnich_JPB12}%
  \BibitemOpen
  \bibfield  {author} {\bibinfo {author} {\bibfnamefont {L.}~\bibnamefont
  {Salasnich}}\ and\ \bibinfo {author} {\bibfnamefont {B.~A.}\ \bibnamefont
  {Malomed}},\ }\href {\doibase 10.1088/0953-4075/45/5/055302} {\bibfield
  {journal} {\bibinfo  {journal} {J. Phys. B At. Mol. Opt. Phys.}\ }\textbf
  {\bibinfo {volume} {45}},\ \bibinfo {pages} {55302} (\bibinfo {year}
  {2012})}\BibitemShut {NoStop}%
\bibitem [{\citenamefont {dos Santos}\ and\ \citenamefont
  {Cardoso}(2019)}]{dosSantos_PLA19}%
  \BibitemOpen
  \bibfield  {author} {\bibinfo {author} {\bibfnamefont {M.~C.}\ \bibnamefont
  {dos Santos}}\ and\ \bibinfo {author} {\bibfnamefont {W.~B.}\ \bibnamefont
  {Cardoso}},\ }\href {\doibase 10.1016/j.physleta.2019.01.064} {\bibfield
  {journal} {\bibinfo  {journal} {Phys. Lett. A}\ }\textbf {\bibinfo {volume}
  {383}},\ \bibinfo {pages} {1435} (\bibinfo {year} {2019})}\BibitemShut
  {NoStop}%
\bibitem [{\citenamefont {dos Santos}\ \emph {et~al.}(2020)\citenamefont {dos
  Santos}, \citenamefont {Malomed},\ and\ \citenamefont
  {Cardoso}}]{Santos_PRE20}%
  \BibitemOpen
  \bibfield  {author} {\bibinfo {author} {\bibfnamefont {M.~C.~P.}\
  \bibnamefont {dos Santos}}, \bibinfo {author} {\bibfnamefont {B.~A.}\
  \bibnamefont {Malomed}}, \ and\ \bibinfo {author} {\bibfnamefont {W.~B.}\
  \bibnamefont {Cardoso}},\ }\href {\doibase 10.1103/PhysRevE.102.042209}
  {\bibfield  {journal} {\bibinfo  {journal} {Phys. Rev. E}\ }\textbf {\bibinfo
  {volume} {102}},\ \bibinfo {pages} {42209} (\bibinfo {year}
  {2020})}\BibitemShut {NoStop}%
\bibitem [{\citenamefont {dos Santos}\ \emph {et~al.}(2019)\citenamefont {dos
  Santos}, \citenamefont {Malomed},\ and\ \citenamefont
  {Cardoso}}]{Santos_JPB19}%
  \BibitemOpen
  \bibfield  {author} {\bibinfo {author} {\bibfnamefont {M.~C.~P.}\
  \bibnamefont {dos Santos}}, \bibinfo {author} {\bibfnamefont {B.~A.}\
  \bibnamefont {Malomed}}, \ and\ \bibinfo {author} {\bibfnamefont {W.~B.}\
  \bibnamefont {Cardoso}},\ }\href {\doibase 10.1088/1361-6455/ab4fb7}
  {\bibfield  {journal} {\bibinfo  {journal} {J. Phys. B At. Mol. Opt. Phys.}\
  }\textbf {\bibinfo {volume} {52}},\ \bibinfo {pages} {245301} (\bibinfo
  {year} {2019})}\BibitemShut {NoStop}%
\bibitem [{\citenamefont {dos Santos}\ \emph {et~al.}(2021)\citenamefont {dos
  Santos}, \citenamefont {Cardoso},\ and\ \citenamefont
  {Malomed}}]{dosSANTOS_EPJ21}%
  \BibitemOpen
  \bibfield  {author} {\bibinfo {author} {\bibfnamefont {M.~C.~P.}\
  \bibnamefont {dos Santos}}, \bibinfo {author} {\bibfnamefont {W.~B.}\
  \bibnamefont {Cardoso}}, \ and\ \bibinfo {author} {\bibfnamefont {B.~A.}\
  \bibnamefont {Malomed}},\ }\href {\doibase 10.1140/epjs/s11734-021-00351-2}
  {\bibfield  {journal} {\bibinfo  {journal} {Eur. Phys. J. Spec. Top.}\ }
  (\bibinfo {year} {2021}),\ 10.1140/epjs/s11734-021-00351-2}\BibitemShut
  {NoStop}%
\bibitem [{\citenamefont {Sakellari}\ \emph {et~al.}(2004)\citenamefont
  {Sakellari}, \citenamefont {Proukakis},\ and\ \citenamefont
  {Adams}}]{Sakellari_JPB04}%
  \BibitemOpen
  \bibfield  {author} {\bibinfo {author} {\bibfnamefont {E.}~\bibnamefont
  {Sakellari}}, \bibinfo {author} {\bibfnamefont {N.~P.}\ \bibnamefont
  {Proukakis}}, \ and\ \bibinfo {author} {\bibfnamefont {C.~S.}\ \bibnamefont
  {Adams}},\ }\href {\doibase 10.1088/0953-4075/37/18/009} {\bibfield
  {journal} {\bibinfo  {journal} {J. Phys. B At. Mol. Opt. Phys.}\ }\textbf
  {\bibinfo {volume} {37}},\ \bibinfo {pages} {3681} (\bibinfo {year}
  {2004})}\BibitemShut {NoStop}%
\bibitem [{\citenamefont {Pethick}\ and\ \citenamefont
  {Smith}(2008)}]{Pethick_08}%
  \BibitemOpen
  \bibfield  {author} {\bibinfo {author} {\bibfnamefont {C.~J.}\ \bibnamefont
  {Pethick}}\ and\ \bibinfo {author} {\bibfnamefont {H.}~\bibnamefont
  {Smith}},\ }\href {\doibase 10.1017/CBO9780511802850} {\emph {\bibinfo
  {title} {{Bose-Einstein Condensation in Dilute Gases}}}}\ (\bibinfo
  {publisher} {Cambridge University Press},\ \bibinfo {address} {Cambridge},\
  \bibinfo {year} {2008})\BibitemShut {NoStop}%
\bibitem [{\citenamefont {Dalfovo}\ \emph {et~al.}(1999)\citenamefont
  {Dalfovo}, \citenamefont {Giorgini}, \citenamefont {Pitaevskii},\ and\
  \citenamefont {Stringari}}]{DALFOVO_RMP99}%
  \BibitemOpen
  \bibfield  {author} {\bibinfo {author} {\bibfnamefont {F.}~\bibnamefont
  {Dalfovo}}, \bibinfo {author} {\bibfnamefont {S.}~\bibnamefont {Giorgini}},
  \bibinfo {author} {\bibfnamefont {L.~P.}\ \bibnamefont {Pitaevskii}}, \ and\
  \bibinfo {author} {\bibfnamefont {S.}~\bibnamefont {Stringari}},\ }\href
  {\doibase 10.1103/RevModPhys.71.463} {\bibfield  {journal} {\bibinfo
  {journal} {Rev. Mod. Phys.}\ }\textbf {\bibinfo {volume} {71}},\ \bibinfo
  {pages} {463} (\bibinfo {year} {1999})}\BibitemShut {NoStop}%
\end{thebibliography}%

\end{document}